\documentstyle[NATO,numreferences,epsf,hyperref]{Crckapb}

\begin{opening}
\title{Damping of collective modes and quasiparticles
in $\lowercase{d}$-wave superconductors}
\subtitle{}

\author{Subir Sachdev and Matthias Vojta}
\institute{Department of Physics\\
Yale University\\
P.O. Box 208120\\
New Haven CT 06520-8120\\
U.S.A.}

\end{opening}

\runningtitle{Collective mode and quasiparticle damping}

\begin{document}


\begin{abstract}
The two-dimensional $d$-wave
superconducting state of the high temperature
superconductors has a number of different elementary excitations:
the spin-singlet Cooper pairs, the spin $S=1/2$ fermionic
quasiparticles, and a bosonic $S=1$ resonant collective mode,
$\phi_{\alpha}$,
at the antiferromagnetic wavevector. Although the $\phi_{\alpha}$
quanta are
strongly coupled to the gapped quasiparticles near the $(\pi,0)$,
$(0,\pi)$ wavevectors (the ``hot spots''), they are essentially
decoupled from the low energy quasiparticles near the nodes of the
superconducting gap. Consequently, distinct and independent low
energy quantum field theories can be constructed for the
$\phi_{\alpha}$ and nodal quasiparticle excitations. We review
recent work introducing a 2+1 dimensional boundary conformal field
theory for the damping of the $\phi_{\alpha}$ excitations by
non-magnetic impurities, which is built on the proximity
to a magnetic ordering transition at which the $\phi_{\alpha}$
condense; the results are compared with neutron scattering
experiments.
Photoemission and THz conductivity
measurements indicate that the nodal quasiparticles undergo strong
inelastic scattering at low temperatures; we propose that this is due
to fluctuations near a quantum phase transition, and critically
analyze candidate order parameters and field theories.
\end{abstract}

\begin{center}
Lectures at the \\
NATO Advanced Study Institute/EC Summer School on \\
{\em New
Theoretical Approaches to Strongly Correlated Systems},\\
Isaac Newton Institute for Mathematical Sciences, \\
Cambridge, UK, April 10-20, 2000.\\
Transparencies at
\href{http://pantheon.yale.edu/~subir}{http://pantheon.yale.edu/\~\/subir}
\end{center}

\section{Introduction}
\label{sec:intro}

The description of high temperature superconductivity in the cuprate
compounds has been a central problem at the frontier of quantum
many body theory in the last decade.
Although many anomalous properties have been observed in the
normal state, both in the over-doped and under-doped regions, no
theoretical consensus has emerged on their origin. Part of the
difficulty is that there appear to be many competing instabilities
and excitations
as one cools down from high temperatures ($T$), and they
are all strongly coupled to each other at intermediate $T$.

However, simplifications do occur at temperatures $T< T_c$, the
critical temperature below which there is an onset of $d$-wave
superconductivity. In this review we shall argue, on the basis of recent
experimental observations, that there is an
important decoupling between different sectors of the excitation
spectrum which carry a non-zero spin, and that
this decoupling allows development of tractable quantum field
theories of the low energy excitations \cite{vs,sbv,vbs,vzs}.
We will make quantitative
predictions for the impurity-induced and intrinsic damping of
these excitations and compare them to experimental results.

\begin{figure}
\epsfxsize=5.1in
\centerline{\epsffile{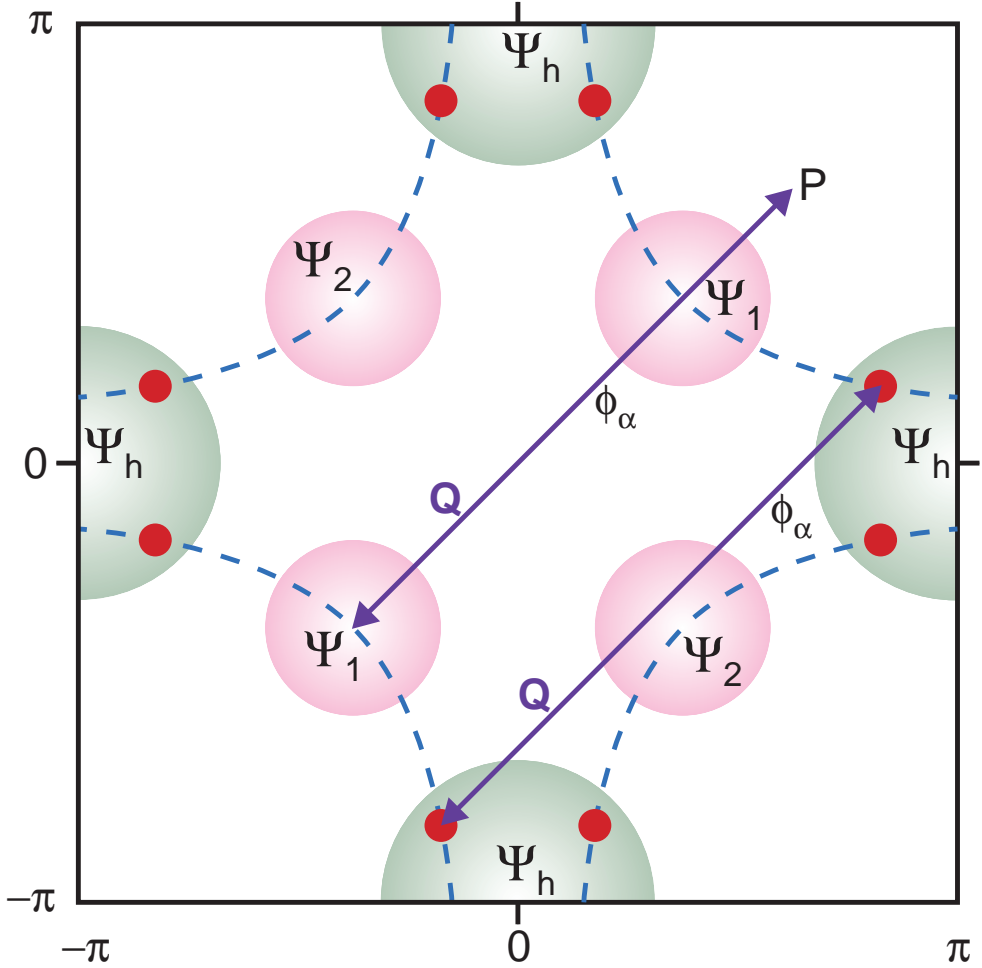}}
\caption{Brillouin zone of the high temperature superconductors at
optimal doping (see {\em e.g.} Ref.~\protect\cite{AbCh}). The dashed
line is the location of the incipient Fermi surface at
intermediate temperatures: the ground state is a $d$-wave
superconductor,
and not a Fermi liquid,
and so there is no sharply defined Fermi surface as $T \rightarrow
0$---the line is merely the location of smooth crossover in the
momentum distribution function. The fermionic, $S=1/2$, quasiparticles
$\Psi_{1,2}$ lie near the nodal points $(\pm K, \pm K)$ (with $K
\approx 0.39 \pi$ \protect\cite{valla})
at which their excitation energy vanishes. The
$\Psi_h$ quasiparticles require an energy $\approx \Delta$ for
their excitation and lie in the vicinity of the ``Fermi surface''
points $(\pm 0.18\pi, \pm \pi)$ and $(\pm \pi, \pm 0.18 \pi)$.
The double-headed arrow at wavevector ${\bf Q} = (\pi,\pi)$
represents the bosonic, $S=1$, resonant collective mode
$\phi_{\alpha}$ scattering fermions between two points in the
Brillouin zone. Notice that the $\Psi_{1,2}$ fermions are
decoupled from the $\phi_{\alpha}$ quanta. In contrast, the $\Psi_h$
fermions couple strongly to the $\phi_{\alpha}$, especially in the
vicinity of the ``hot spots'' denoted by the small filled circles.
 }
\label{fig1}
\end{figure}

Let us list the elementary excitations of the $d$-wave
superconductor and nearby phases:
\paragraph{(A) \underline{Cooper Pairs:}}
The superconductivity is of course a consequence
of the condensation of spin $S=0$, charge $2e$ Cooper pairs. Below
$T_c$, the excitations of the phase of the condensate are
responsible for the superflow, and for the plasmon excitations. In
this paper, we will be primarily concerned with the damping of
excitations which carry spin, and these couple only weakly to the
phase excitations in a well-formed superconductor at low $T$: so
we will neglect the phase excitations in the body of the paper.
These phase excitations become more important near a $T=0$
superconducting-insulator transition, but we will not consider
such a situation here. Above $T_c$, phase
fluctuations \cite{uemura,doniach,nandini,emery,tvr} are surely
important for the transport properties, and they also couple
strongly to some of the fermionic quasiparticle excitations:
we will briefly discuss this phenomenon further below.
\paragraph{(B) \underline{$S=1/2$ fermionic quasiparticles:}}
These are the familiar Bogoliubov
quasiparticles in a BCS theory of the superconducting state.
Because of the $d$-wave symmetry of the order parameter, their
energies vanishes at four nodal points in the Brillouin zone -
$(\pm K, \pm K)$, with $K= 0.391 \pi$ for optimally doped ${\rm
Bi}_2 {\rm Sr}_2 {\rm Ca Cu}_2 {\rm O}_{8+\delta}$ \cite{valla}. We will denote
the fermionic excitations in the vicinity of these points by the
Nambu spinors $\Psi_{1,2}$ (see Fig~\ref{fig1} and further details
in Section~\ref{damp}).
It is also interesting to consider the fermionic
excitations near the $(\pi, 0)$ and $(0, \pi)$ points: here the
pairing amplitude has its largest value and so there is a large
energy gap, $\Delta$, towards exciting the quasiparticles. We will denote
these high energy quasiparticles generically by $\Psi_h$ (see
Fig~\ref{fig1}). Because of their large pairing amplitude, the
$\Psi_h$ quasiparticles couple efficiently to the phase
fluctuations discussed above in (A), and are expected to have a
rapidly decreasing lifetime once the phase fluctuations
proliferate above $T_c$ \cite{gls}. In contrast, the nodal quasiparticles,
$\Psi_{1,2}$, are in a region of vanishing pairing, and are
essentially decoupled from the phase fluctuations: as we will
discuss below, other mechanisms will be required to damp the
$\Psi_{1,2}$ quasiparticles.
\paragraph{(C) \underline{$S=1$ resonant collective mode:}}
Neutron scattering experiments
observe a sharp resonance peak at an energy $\Delta_{\rm res}$
in the scattering cross section at
the antiferromagnetic wavevector,
${\bf Q}$ \cite{rossat1,mook,tony3,bourges,he}. We will view this
bosonic $S=1$ resonant collective mode, $\phi_{\alpha}$
($\alpha=x,y,z$ are the spin components), as that expected in a
paramagnetic phase across a magnetic disordering quantum phase
transition \cite{CSY}. (For the case where ${\bf Q} = (\pi,\pi)$,
the $\phi_{\alpha}$ are real, while for incommensurate ${\bf Q}$,
the $\phi_{\alpha}$ become complex; we will explicitly treat the
commensurate case here, although the generalization to the
incommensurate case is straightforward \cite{vbs}, and does not
modify any of the scaling arguments, including the central result
(\ref{gamma}).) Our identification of $\phi_{\alpha}$
is similar to the view that it is a
$S=1$ particle-hole bound state in a $d$-wave
superconductor \cite{levin,morr,brinck}.
However, as we shall discuss in more detail below, appealing to
the proximity of a quantum phase transition allows a systematic
treatment of the strongly relevant self-interactions of the
$\phi_{\alpha}$ field. Such a collective mode has also been
discussed in models with a special SO(5) symmetry \cite{zhang},
but we shall
not appeal to symmetries beyond the usual SU(2) spin symmetry in
our treatment. The coupling of the $\phi_{\alpha}$ to the
fermionic quasiparticles in (B) is illustrated in
Fig~\ref{fig1}. Momentum conservation allows a strong coupling
between the $\Psi_h$ and the $\phi_{\alpha}$ in the vicinity of
the so-called ``hot spots'' \cite{AbCh,mopi,AbCh1,MoCh}.
This coupling leads to strong mutual
damping of $\Psi_h$ and $\phi_{\alpha}$ above $T_c$. Below
$T_c$, the same coupling is surely an important ingredient in the pairing of the
$\Psi_h$ fermions \cite{scalapino}: this conclusion is supported by the
``shake-off'' satellite peaks, separated by the
$\Delta_{\rm res} \sim$ 40 meV, observed in the
photoemission and optical spectra \cite{AbCh1,carbotte,munzar}.
A key ingredient in our
discussion is also clear from Fig~\ref{fig1}: the coupling between
the low energy $\Psi_{1,2}$ fermions and the $\phi_{\alpha}$ mode
is strongly suppressed by momentum conservation---moving a distance
${\bf Q}$ from a nodal point places one in a section of the
Brillouin zone ({\em e.g.\/} near the point $P$ in Fig~\ref{fig1})
where the fermionic excitations cost over 100 meV.
Remnants of the shake-off peaks just noted are also seen along the
$(1,\pm 1)$ directions\cite{kami} not too far from the point $P$, but this
does not qualitatively modify the low energy $\Psi_{1,2}$
excitations within the shaded circles.

We have now collected all the ingredients necessary to motivate
our recent computations.

In Section~\ref{qimp} we will discuss the $T=0$ broadening of the
$\phi_{\alpha}$ collective mode by the substitution of a small concentration
of non-magnetic impurities like Zn or Li on the Cu sites, and compare
our theoretical results \cite{sbv,vbs} to experimental observations \cite{fong}.
It should be clear from the discussion above that, in the absence of
such extrinsic broadening, it is possible for the $\phi_{\alpha}$
resonance to be infinitely sharp at $T=0$ in a $d$-wave
superconductor. The $\phi_{\alpha}$ quanta couple strongly to the
$\Psi_h$ fermions, but these induce no damping as long as
$\Delta_{\rm res} < 2 \Delta$
($\Delta\approx 40$ meV near optimal doping \cite{mesot}).
Along the diagonals of the Brillouin
zone, momentum conservation prohibits coupling to the gapless
nodal fermions $\Psi_{1,2}$, and only allows couplings to
quasiparticle excitations whose excitation energy exceeds
$\Delta_{\rm res}$.

Section~\ref{damp} will consider {\em intrinsic} $T$-dependent
damping of the nodal fermionic quasiparticles by inelastic
scattering \cite{vs,vzs}. We have already discussed the damping of the gapped
$\Psi_h$ above. Below $T_c$, photoemission
experiments \cite{shen,campu,fed,mesot,mesot2} observe
negligible damping of the $\Psi_h$ fermions, and this is
consistent with our considerations: the phase fluctuations in (A)
are suppressed, while the $\phi_{\alpha}$ mode in (C) leads to coherent
pairing of $\Psi_h$ fermions. Above $T_c$, phase fluctuations
proliferate, and their strong coupling to the $\Psi_h$ is expected
to lead to significant inelastic scattering of the $\Psi_h$ \cite{gls};
additional scattering is also expected from the ``hot-spot''
coupling to the $\phi_{\alpha}$, and these expectations are
consistent with experimental observations. However neither of the
fluctuations in (A) or (C) couple to the nodal fermions
$\Psi_{1,2}$: so, with our present considerations we would
conclude that the nodal fermions should be very sharp both below
and above $T_c$. The actual experimental situation is dramatically
different \cite{valla,corson}---these nodal
quasiparticles have a large inverse lifetime, which
decreases roughly linearly with $T$,
and an imaginary component of a self energy which is
roughly linearly proportional to frequency, $\omega$,
for $\hbar \omega > k_B T$.
Moreover, these damping rates change
smoothly through $T_c$, with little sign of the superconducting
transition. We have to appeal to other inelastic damping mechanisms to
explain these observations, and we will present a critical classification
of candidates in Section~\ref{damp}.

\section{Impurities and the $S=1$ resonant collective mode}
\label{qimp}

\begin{figure}
\epsfxsize=5in
\centerline{\epsffile{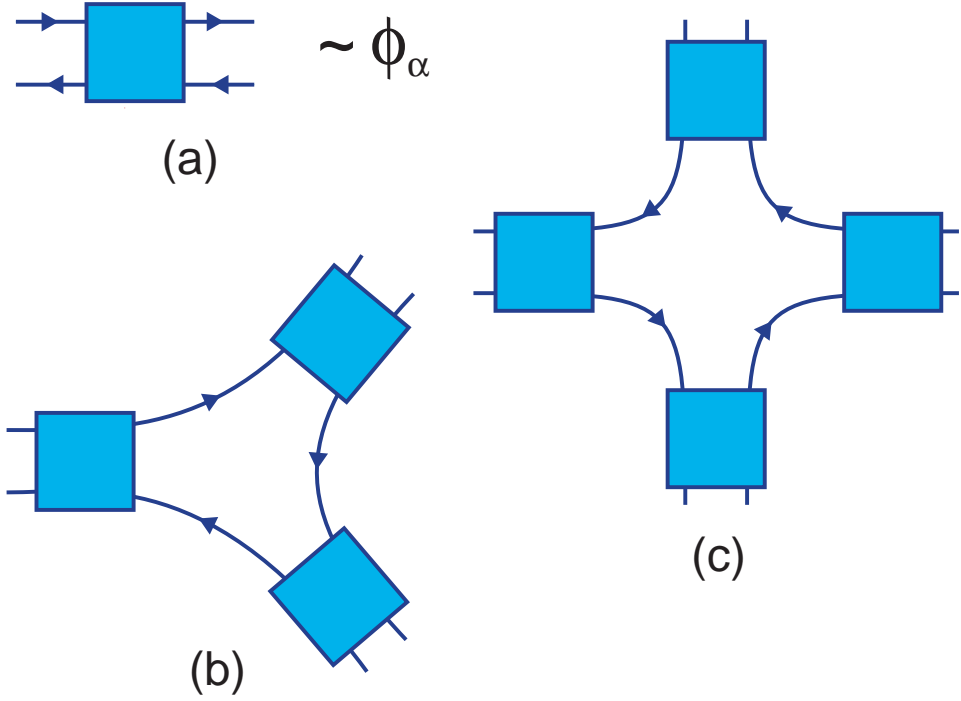}}
\caption{
(a) Particle-hole propagator which has a $S=1$ bound state at the
antiferromagnetic wavevector representing the $\phi_{\alpha}$
quanta. (b) and (c) Three- and four-point self interactions of the
spin excitations.
}
\label{fig2}
\end{figure}

Before discussing the effect of impurities, we state our model for
the spin collective mode, $\phi_{\alpha}$, in the clean $d$-wave superconductor. A
popular approach in recent work \cite{levin,morr,brinck}
has been to compute the dynamic spin
susceptibility of the underlying electrons and to identify this mode as a
bound state pole near the antiferromagnetic wavevector: this is
schematically indicated in Fig~\ref{fig2}a.
However all such
studies have so far neglected three- and four-point (and higher multi-point)
self interactions of the spin excitations, which are
schematically indicated in Figs~\ref{fig2}b and~\ref{fig2}c
respectively. One of the key points of our work is that it is
essential to include the interactions in Figs~\ref{fig2}b and~\ref{fig2}c
in the low energy theory: these are strongly relevant
perturbations, and in a sense, their effective strength is
infinite (higher multi-point interactions can, however, be neglected).
The simple results for impurity-induced damping we shall
quote below rely on hyperscaling properties, and these are a direct
consequence of the self interactions in Fig~\ref{fig2}.

So how does one obtain a tractable theory which includes the
interactions in Fig~\ref{fig2} ?
Our strategy is to appeal to the proximity of a
$T=0$ magnetic ordering transition at which the excitation energy
of $\phi_{\alpha}$, $\Delta_{\rm res}$, vanishes.
This is a quantum
phase transition driven by the condensation of $\phi_{\alpha}$ to
a state with coexisting superconductivity and collinear
spin density wave order.
We will develop a theory for the quantum-critical point of this
transition, and then use powerful field-theoretic methods to
expand back into the region where $\Delta_{\rm res}$ is non-zero.
The values of the interactions in Fig~\ref{fig2} will be
universally determined by the underlying structure of the
expansion in relevant perturbations of the critical field theory.
Such an approach effectively reduces to an expansion in
$\Delta_{\rm res}/J$, where $J$ is a microscopic exchange constant,
and the smallness of this ratio is our primary assumption.

We shall use a theoretical model (which is supported by the
`pseudo-gap' phenomenology) in which the magnetic ordering
transition occurs while
the excitation energy of the $\Psi_h$ remains non-zero, and so
these fermions can be neglected in the critical theory of the
transition. As discussed above, the $\Psi_{1,2}$ fermions have
little coupling to the magnetic excitations, and so an
action for the transition can be expressed in terms of
the $\phi_{\alpha}$ alone. By analogy with theories developed for
the magnetic transition in insulators, and using general symmetry
arguments \cite{book}, we can write down the following effective action for
the bulk $\phi_{\alpha}$ fluctuations in the $d$-wave
superconductor:
\begin{equation}
{\cal S}_{b} = \int d^2 x \int d \tau \bigg[\frac{1}{2} \left(
(\partial_{\tau} \phi_{\alpha})^2 + c^2 ( \nabla_{x}
\phi_{\alpha} )^2  + s \phi_{\alpha}^2
\right) + \frac{g_0}{4!} \left( \phi_{\alpha}^2 \right)^2 \bigg];
\label{sb}
\end{equation}
the fermionic excitations with spin, $\Psi_{1,2}$, $\Psi_h$ have
been integrated out, and for reasons already discussed, serve only
to renormalize the values of the couplings in (\ref{sb}).
The parameter $c$ is the velocity of spin-waves in the ordered phase,
and $s$ tunes the system between the two phases which lie on either
side of a critical value $s=s_c$. The quartic
non-linearity, $g_0$, corresponds to the interaction in
Fig~\ref{fig2}c. How about the cubic coupling in Fig~\ref{fig2}b ?
This is implicitly accounted for in (\ref{sb}) in a manner we now
describe. By momentum conservation, if two of the particle-hole
propagators in Fig~\ref{fig2}b carry momentum ${\bf Q}$, the third
must carry momentum $2 {\bf Q} \approx 0$, {\em i.e.}, it represents
the ferromagnetic spin component $L_{\alpha}$. This has allowed
three-point couplings with the $\phi_{\alpha}$, including the
kinematic term \cite{book} in the action $\sim i \epsilon_{\alpha\beta\gamma}
L_{\alpha} \phi_{\beta} \partial_{\tau} \phi_{\gamma}$.
The $L_{\alpha}$ fluctuations are not critical, and after
integrating them out, one obtains renormalizations of terms
already present in (\ref{sb}) \cite{book}.

The magnetic properties of the bulk quantum phase transition
described by of ${\cal S}_b$
have been worked out in some detail \cite{CHN,CSY}, and many aspects are in
agreement with trends in NMR and neutron scattering experiments on
the high temperature superconductors \cite{CS,sciencereview}.
Here, we will only need a few well-known scaling properties of the critical
point of (\ref{sb}). Upon interpreting $\tau$
as a third spatial dimension, ${\cal S}_b$ can also represent the
partition function of a classical Heisenberg ferromagnet in dimension $D=3$
at finite
temperature, and its Curie transition corresponds to the quantum-critical
point we are interested in. Both occur at a critical
value $r=r_c$, where there the functional integral over $\phi_{\alpha}$
is invariant under the scale transformation
\begin{eqnarray}
x & \rightarrow & x/b \nonumber \\
\tau & \rightarrow & \tau/b \nonumber \\
\phi_{\alpha} & \rightarrow &  b^{(1+\eta_H)/2} \phi_{\alpha} \nonumber \\
\phi_{\alpha}^2 & \rightarrow &  b^{3-1/\nu_H} \phi_{\alpha}^2 ,
\label{scaleH}
\end{eqnarray}
where $b$ is a rescaling factor, and
$\nu_H$ and $\eta_H$ are known critical exponents of the $D=3$
classical Heisenberg model. The last transformation in
(\ref{scaleH}) represents the mapping of the composite operator
$\phi_{\alpha}^2$, and its scaling dimension is not simply twice
that of $\phi_{\alpha}$ because of corrections due to the $g_0$
interaction in (\ref{sb}).

We now turn to the effect of a dilute concentration of impurities.
We will outline the central ingredients leading to our main
result, and refer the reader to Ref.~\cite{vbs} for further
details. Consider a single impurity at $x=0$; by ``impurity'' we
mean an {\em arbitrary} localized deformation in the vicinity of
$x=0$. One consequence of any such deformation will be a change in
the value of $s$ near $x=0$, and this will lead to the following
term in the action
\begin{equation}
\zeta \int d \tau \phi_{\alpha}^2 (x=0,\tau).
\label{zeta}
\end{equation}
Under the scale transformation (\ref{scaleH}), we see immediately
that $\zeta$ has scaling dimension $1/\nu_H - 2 \approx -0.57 $,
and is therefore {\em irrelevant} at the critical point of (\ref{sb}), and it
has only weak effects on the bulk properties. To obtain a
local, relevant, perturbation on the bulk fluctuations, we need to
consider quantum mechanical effects associated with Berry
phases. The Berry phases accumulated by the precession of spins in
the host antiferromagnet cancel almost completely upon an average
over the lattice sites \cite{book}; however, in the presence of
impurities it is entirely possible that this cancellation is
disrupted, and a residual Berry phase of spin $S$ ($S$ must be an
integer or half-odd-integer) survives \cite{SY,sigrist1}. To
account for this Berry phase we introduce a single unit vector
$n_{\alpha} (\tau)$ ($n_{\alpha}^2 (\tau) = 1$) representing the orientation of
the net uncompensated spin,  and the action
\begin{equation}
{\cal S}_{\rm imp} =  i S\int d \tau
A_{\alpha} (n) \frac{d n_{\alpha}(\tau)}{d \tau},
\label{simp}
\end{equation}
where $A_{\alpha}$ is a function of $n_{\alpha}$ defined by
$\epsilon_{\alpha\beta\gamma} \partial A_{\beta} /\partial
n_{\gamma} = n_{\alpha}$. This Berry phase is intimately connected
to the fact that an external magnetic field will lead to a Curie
susceptibility $= S(S+1)/3T$ from the impurity spin in the
non-magnetic phase (in the absence of Kondo screening--see below);
this response is divergent as $T \rightarrow 0$ and is a
reflection of a $(2S+1)$-fold degenerate level near the
impurity. For the case of a non-magnetic Zn or Li ion
replacing a magnetic $S=1/2$ Cu ion in the high temperature
superconductors, the above arguments on Berry phases strongly
suggest that each such impurity should contribute a term like
(\ref{simp}) with $S=1/2$. This conclusion is supported
by NMR experiments \cite{alloul,alloul2a,julien},
and we will assume its validity in our
discussion below. Other authors have modeled the Zn ion solely
as a non-magnetic scatterer in the unitarity
limit \cite{sasha1,kallin,fulde,pepin,ting}. In such a
model, the fermionic quasiparticles form
quasi-bound states at the impurity sites at the Fermi level; we believe that
after accounting for the strong local Coulomb repulsion at the impurity
site, each
bound state will capture only a single electron, and the low
energy physics will then be described by (\ref{simp}) with
$S=1/2$ (see also Ref~\cite{fulde}).

We now need to couple the impurity degree of freedom,
$n_{\alpha}$, to those of the host. The most important coupling is
the simple linear term
\begin{equation}
{\cal S}_c = \gamma \int d \tau n_{\alpha} (\tau) \phi_{\alpha}
(x=0,\tau).
\label{sc}
\end{equation}
To compute the scaling dimension of $\gamma$ at the fixed
point where the impurity and bulk degrees of freedom are
decoupled, we note
that if $n_{\alpha} \rightarrow n_{\alpha}$
under the transformation
(\ref{scaleH}), the impurity action (\ref{simp}) remains
invariant. Under such a mapping, $\gamma$ has dimension
$(1-\eta_H)/2 \approx 0.48$.
Unlike $\zeta$, the coupling $\gamma$ is therefore {\em relevant} at the
$\gamma=0$ fixed point, and plays a central role in the main
results presented below. We can also imagine a Kondo coupling,
$J_K$ between the spin, $n_{\alpha} (\tau)$, and the host fermions
$\Psi_{1,2}$, $\Psi_h$. However, the fermionic, single particle
density of states vanishes at the Fermi level, and this
dramatically reduces the possibility of Kondo screening of the
impurity spin: with particle-hole symmetry, there is no
Kondo screening even upto $J_K = \infty$, while without
particle-hole symmetry, the spin is screened only above an
appreciable threshold value of $J_K$
\cite{withoff,CJ,ingersent,bulla}. We will assume that no Kondo
screening has occurred over the experimentally relevant
temperature range.

The above arguments suggest that a theory of the impurity spin
dynamics will emerge from a complete renormalization group
analysis of ${\cal S}_b + {\cal S}_{\rm imp} + {\cal S}_c$.
This has been carried out in Ref~\cite{vbs}, and the final results
are quite simple: the bulk phase transition at $s=s_c$ is the only
critical point, and $(s-s_c)$ remains the only relevant
perturbation at this critical point; both $g$ and $\gamma$
approach fixed point values $g^{\ast}$ and $\gamma^{\ast}$
(related phenomena were noted earlier in simpler
models \cite{SY2,si,Sengupta}).
There is no separate critical point associated with the impurity
degrees of freedom, as is often the case in the theory critical
phenomena on boundaries \cite{cardybook}. A remarkable consequence
of this is that the single energy scale, $\Delta_{\rm res}$, which
characterized the dynamics of the paramagnet in the host system
\cite{book}, is also all that is needed to completely characterize
the dynamics in the vicinity of the impurity. Of course, the
quantized number $S$ in (\ref{simp}) also influences the values of the various
universal scaling functions.

Now consider a dilute concentration of impurities, $n_{\rm imp}$,
each described by the analog of (\ref{simp}) and (\ref{sc}),
placed at random locations in the $d$-wave superconductor. We will
answer the following key question, relevant to the neutron
scattering experiments on Zn doped ${\rm Y} {\rm Ba}_2 {\rm
Cu}_3 {\rm O}_7$ by Fong {\em et al.} \cite{fong}. At what energy
scale, $\Gamma$, does the $S=1$ resonant pole at energy $\Delta_{\rm res}$
get broadened by the impurities ? The implication of the
renormalization group arguments above is that, for small $\Delta_{\rm res}/J$, $\Gamma$ is
universally determined by the only dimensionful parameters
available to us: $n_{\rm imp}$, $\Delta_{\rm res}$, and the
velocity $c$. Making the mild assumption that, for small $n_{\rm
imp}$, $\Gamma$ must be linearly proportional to $n_{\rm imp}$
(this is supported by explicit computations \cite{vbs}), simple
dimensional analysis of the length and time scales allows us to
conclude
\begin{equation}
\Gamma = C_S \frac{(\hbar c)^2}{\Delta_{\rm res}} n_{\rm imp},
\label{gamma}
\end{equation}
where $C_S$ is a universal number. All corrections to
(\ref{gamma}) will be suppressed by positive powers of $\Delta_{\rm
res}/J$. Notice also the inverse dependence on the small energy
scale $\Delta_{\rm res}$: this is an indication of the strong
effect of the relevant coupling in (\ref{sc}). For an impurity
with $S=0$, the simplest allowed coupling is $\zeta$, and its
irrelevance implies $C_0 = 0$, and the corrections just noted will
be the leading contributions. We have estimated $C_{1/2}$ in a
self-consistent non-crossing approximation and obtained $C_{1/2} \approx
1$.

It is useful to rewrite our main result (\ref{gamma}) in a
different manner:
\begin{equation}
\frac{\Gamma}{\Delta_{\rm res}} \sim n_{\rm imp} \xi^2,
\end{equation}
where $\xi = \hbar c/\Delta_{\rm res}$ is a correlation length.
So if we imagine a ``swiss cheese'' model \cite{tomo} where each impurity
makes a hole of radius $\xi$, the inverse $Q$ of the resonance is
of order the fractional volume of holes in the swiss cheese.

The numerical predictions of (\ref{gamma}) are in good agreement
with the observations of Fong {\em et al.} \cite{fong}. We use $n_{\rm imp} =
0.005$, $\Delta_{\rm res} = 40$ meV, and the spin-wave velocity in
the insulator $\hbar c = 0.2$ eV, and obtain $\Gamma = 5$ meV.
This compares well with the observed value of $4.25$ meV.
We have also predicted detailed lineshapes for the
impurity-induced broadening, and these will hopefully be tested in
future, higher precision experiments.

\section{Inelastic damping of the nodal quasiparticles}
\label{damp}

As we noted at the end of Section~\ref{sec:intro}, recent
experimental observations \cite{valla,corson} of a short lifetime $\sim \hbar/k_B
T$, and a large frequency-dependent self energy,
for the nodal quasiparticles both above and below $T_c$, are
puzzling in the light of the very weak coupling between the $\Psi_{1,2}$
and both the $\phi_{\alpha}$ and phase
fluctuations. Consequently, we have to appeal to a separate
decoupled sector of low energy fluctuations to explain this
anomalous damping \cite{vs,vzs}.

\begin{figure}
\epsfxsize=5in
\centerline{\epsffile{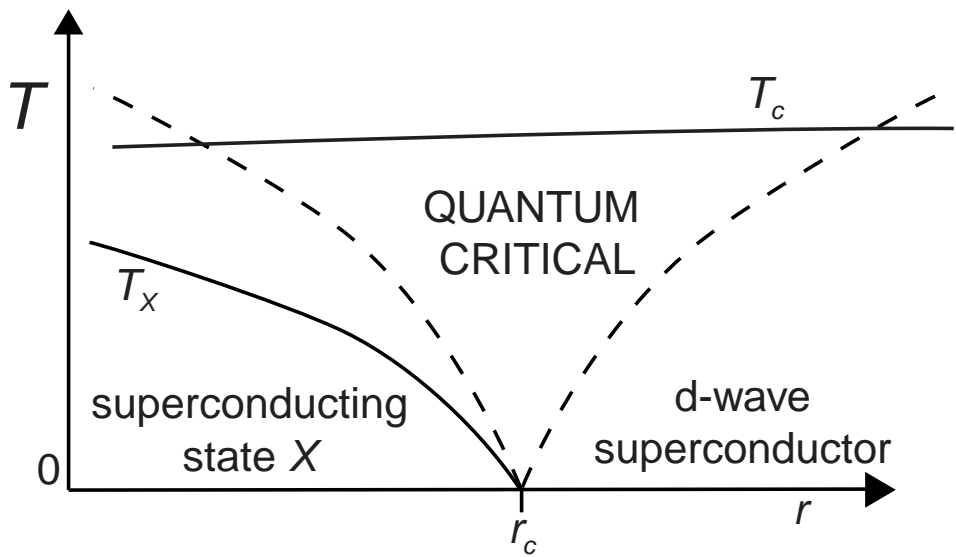}}
\caption{
Finite temperature ($T$) phase diagram in the vicinity of a second
order quantum phase transition from a $d$-wave superconductor as a
function of some parameter in the Hamiltonian,  $r$ (which is possibly, but not
necessarily, the hole concentration $\delta$).
Superconductivity is present at temperatures below $T_c$, and the
superfluid density is non-zero on both sides of $r_{c}$.
The state $X$ is characterized by some other order parameter (in
addition to superconductivity) which vanishes above a temperature
$T_X$. We discuss a number of possibilities for the state $X$
in the text--only one naturally satisfies the requirement of leading
to an inverse lifetime $\sim k_B T/\hbar$ for the nodal quasiparticles
$\Psi_{1,2}$ in the quantum-critical region, and negligible
damping of the $\Psi_h$ quasiparticles below $T_c$:
the $(d_{x^2-y^2} + i d_{xy})$-wave superconductor.
Our computations for the quantum-critical point are carried
out below $T_c$, but the results should also apply above $T_c$
as long as the quantum-critical length $\sim T^{-1/z}$ remains
shorter than the phase coherence length.
We emphasize that we are not requiring the
high temperature superconductors to have $(d_{x^2-y^2} + i d_{xy})$
order in the ground state (although it is permitted): the coupling
$r$ could be larger than $r_c$, but it should be close enough
that the system enters the quantum-critical region at some
low $T$.
}
\label{fig3}
\end{figure}

A natural way of obtaining inverse lifetimes of order $k_B
T/\hbar$, and self energies of order $\omega$,
is to assume
that the system is in the quantum-critical region of a $T=0$
quantum phase transition for which the $\Psi_{1,2}$ fermions are
central critical degrees of freedom \cite{SY,book} (see
Fig~\ref{fig3}); the quantum-critical point should be described by an
interacting quantum field theory below its
upper critical dimension, so that universal low-energy
fluctuations dominate the interactions.
A further constraint is that the critical fluctuations should be
decoupled from the $\Psi_h$ fermions, as these remain
undamped below $T_c$: so the new low energy mode associated with
the onset of state $X$ in Fig~\ref{fig3} couples strongly to the $\Psi_{1,2}$
but not the $\Psi_h$. It will turn out that these constraints are
rather difficult to satisfy, and lead to an essentially unique
identification of the state $X$ in Fig~\ref{fig3}.

Clearly, the
magnetic transition in Section~\ref{qimp} cannot be the required
transition because the $\Psi_{1,2}$ fermions are innocuous
spectators of its critical field theory. We will list below
many of the order parameters that have been considered in
the literature in the last decade, and discuss whether they
satisfy
the requirements we have imposed on the state $X$.

Before we embark on this, let us recall the effective action for
the nodal quasiparticles in the $d$-wave superconductor well away
from the quantum-critical point, $r \gg r_c$.
We denote the
components of the electron annihilation operator, $c_a$, in the vicinity
of the four nodal points $(K,K)$, $(-K,K)$, $(-K,-K)$, $(K, -K)$
by $f_{1a}$, $f_{2a}$, $f_{3a}$, $f_{4a}$ respectively,
where $a= \uparrow, \downarrow$ is the electron spin component.
The 4-component Nambu spinors are $\Psi_1 =
(f_{1a}, \varepsilon_{ab} f_{3b}^{\dagger})$
and  $\Psi_2 =
(f_{2a}, \varepsilon_{ab} f_{4b}^{\dagger})$ where
$\varepsilon_{ab}$ is an antisymmetric tensor with
$\varepsilon_{\uparrow \downarrow} = 1$.
The action is then
\begin{eqnarray}
{\cal S}_{\Psi} &=& \int  \frac{d^d k}{(2 \pi)^d} T  \sum_{\omega_n}
\Psi_1^{\dagger}  \left(
- i \omega_n + v_F k_x \tau^z + v_{\Delta} k_y \tau^x \right) \Psi_1  \nonumber \\
&~&  +\int  \frac{d^d k}{(2 \pi)^d} T  \sum_{\omega_n}
\Psi_2^{\dagger}  \left(
- i \omega_n + v_F k_y \tau^z + v_{\Delta} k_x \tau^x \right) \Psi_2 .
\label{dsid1}
\end{eqnarray}
Here $\tau^{\alpha}$ are Pauli matrices which act in the fermionic
particle-hole space, $k_{x,y}$ measure the wavevector from the nodal points and
have been rotated
by 45 degrees from the axes of the square lattice, and $v_{F}$, $v_{\Delta}$
are velocities.
The action ${\cal S}_{\Psi}$ has a scale invariance which will be
useful in our considerations below
\begin{eqnarray}
x &\rightarrow& x/b \nonumber \\
\tau &\rightarrow& \tau/b \nonumber \\
\Psi_{1,2} (x, \tau) &\rightarrow& b \Psi_{1,2} (x/b,\tau/b),
\label{scale}
\end{eqnarray}
where $b$ is a rescaling factor. We can illustrate the power of
scaling arguments by showing how this transformation allows us to
quickly deduce damping produced by ordinary screened Coulomb
interactions (the screening is performed by the Cooper pairs in the condensate).
These interactions lead to couplings like
\begin{equation}
v \int d^2 x \int d \tau (\Psi_1^{\dagger} \tau^z \Psi_1 )^2 ,
\label{coul}
\end{equation}
and it is easy to see that under (\ref{scale}), $v$ has scaling
dimension -1. Therefore $v$ is {\em irrelevant}, and perturbation
theory in $v$ should be reliable. The fermion self energy,
$\Sigma_f$, acquires an imaginary part at order $v^2$; using the
fact that under (\ref{scale}) the frequency and momentum dependent
self energy transforms as $\Sigma_f \rightarrow b \Sigma_f$, the
result (\ref{coul}) immediately leads to $\mbox{Im} \Sigma_f \sim v^2 T^3$
(and similarly for the $\omega$ dependence). This damping rate
appears far too weak to explain the experimental observations.

We now turn to considerations of quantum critical points, and
list various plausible candidates for the state $X$ in
Fig~\ref{fig3}. We will only consider order parameters for $X$
which break simple underlying symmetries of the Hamiltonian:
the symmetry of the square lattice space group, time-reversal,
and spin rotation (the last we have already discussed above).
More complicated transitions with non-local order parameters and
deconfinement transitions are also possible, but we will not
consider them here \cite{senthil}.

\paragraph{\underline{Staggered flux phase:}}
Many investigators \cite{affmar,schulzoaf,biq,wang,nerses,nayak,ivanov}
have considered the possibility of $d$-wave
superconductivity coexisting with a staggered distribution of
orbital currents (or an ``orbital antiferromagnet''). This state
$X$
is characterized by the expectation value \cite{nayak}
\begin{equation}
\langle c_{k+{\bf Q},a}^{\dagger} c_{k,a} \rangle = i \phi
(\cos k_x - \cos k_y),
\label{oaf1}
\end{equation}
where, as before, ${\bf Q} = (\pi,\pi)$, and $\phi$ is a real
order parameter. Momentarily neglecting the fermionic excitations,
we can, just as for (\ref{sb}), write down an effective action for
$\phi$ fluctuations purely on symmetry grounds:
\begin{equation}
{\cal S}_{\phi} = \int d^2 x \int d \tau \bigg[\frac{1}{2} \left(
(\partial_{\tau} \phi )^2 + c^2 ( \nabla_{x}
\phi )^2  + r \phi^2
\right) + \frac{\tilde{g}}{4!} \phi^4 \bigg];
\label{sI}
\end{equation}
(without strict particle-hole symmetry, a first order time
derivative term is potentially allowed, but the only possible relevant term,
$\phi \partial_\tau \phi$, is a total derivative $(1/2) \partial_{\tau}
\phi^2$).
As is well known, ${\cal S}_{\phi}$ describes the phase transition in
the Ising model in $D=3$ spacetime dimensions. At the critical
point $r=r_c$, ${\cal S}_b$ is also invariant under the
extension of (\ref{scale}) to the
analog of the scale transformations in (\ref{scaleH})
\begin{eqnarray}
\phi & \rightarrow &  b^{(1+\eta_I)/2} \phi \nonumber \\
\phi^2 & \rightarrow &  b^{3-1/\nu_I} \phi^2 ,
\label{scaleI}
\end{eqnarray}
where now $\nu_I$ and $\eta_I$ are the exponents of the $D=3$
Ising model. To decide if this purely Ising description of the
transition is correct, we have to test its stability to a coupling
between the $\phi$ and the $\Psi_{1,2}$. From (\ref{oaf1}) we see
that $\phi$ carries momentum ${\bf Q}$, and so the momentum
conservation constraints upon its coupling to the fermions are
identical to those of the spin mode $\phi_{\alpha}$ in
Fig~\ref{fig1}: unless $K=\pi/2$, there is no linear coupling
to the fermionic excitations which is linear in $\phi$. For general $K$, the simplest
allowed coupling is
\begin{equation}
w \int d^2 x \int d \tau \phi^2 \Psi_1^{\dagger}
\tau^z \Psi_1,
\label{defw}
\end{equation}
and similarly for $\Psi_2$. Using (\ref{scale}), (\ref{scaleI}),
we deduce that the scaling dimensions of $w$ is $1/\nu_I -2
\approx -0.41$: consequently $w$ is {\em irrelevant}, and the
critical theory for the transition is (\ref{sI}) alone. The
fermions $\Psi_{1,2}$ are not part of the critical theory, and
their inverse lifetimes can be estimated by perturbation theory in
$w$.
Using the scaling dimension of $w$ above, we deduce that the
fermionic self energy
has the
following $T$ dependence in the quantum-critical region of
Fig~\ref{fig3}: $\mbox{Im} \Sigma_f \sim w^2 T^{5-2/\nu_I} \approx w^2
T^{1.83}$ (and $\mbox{Im} \Sigma_f \sim w^2 \omega^{1.83}$ for
$\hbar \omega > k_B T$). This is a super-linear power, which does not appear
compatible with experimental observations.
Finally, we note that the special case $K=\pi/2$ has also been
considered in Ref~\cite{vzs}: then a coupling term linear in $\phi$
is also allowed \cite{nayak}, but its contribution to $\Sigma_f$
vanishes with an even higher power of $T$.

\paragraph{\underline{Charge Stripes:}}
We consider the onset of a charge density wave in a $d$-wave
superconductor, a transition to a state $X$
defined by the order parameters
\begin{equation}
\langle c_{k+{\bf G}_x,a}^{\dagger} c_{k,a} \rangle =
\Phi_x~~~;~~~
\langle c_{k+{\bf G}_y,a}^{\dagger} c_{k,a} \rangle = \Phi_y
\label{cdw}
\end{equation}
where ${\bf G}_x = (G,0)$, ${\bf G}_y = (0,G)$ are the ordering
wavevectors, and $\Phi_{x,y}$ are complex order parameters.
Again, constraints from momentum conservation are
rather severe. Unless $G=2K$, there is no coupling between the
$\Phi_{x,y}$ and the $\Psi_{1,2}$ fermions. Existing experimental
observations of charge stripe formation easily satisfy $G \neq 2
K$. Under these conditions, the damping of the $\Psi_{1,2}$ from
the critical charge fluctuations can be estimated as in the
staggered-flux case above: the simplest allowed couplings,
as in (\ref{defw}), are $\sim |\Phi_x|^2 \Psi_1^{\dagger} \tau^z
\Psi_1$ etc., and we obtain $\mbox{Im} \Sigma_f
\sim ({\rm max}(\omega, T))^{5-2/\nu}$.
Any reasonable model of the critical theory of the $\Phi_{x,y}$
\cite{vzs} has $\nu > \nu_I$, and so the fermion damping is
rather weak. The special case $G=2K$ has also been analyzed in
Refs~\cite{vs,vzs}: it does yield fermion damping compatible with
experimental observations \cite{valla,corson}, but, as we have
already noted, this mode-locking of the charge stripe
and fermionic nodal wavevectors is not supported by experiments.

\paragraph{\underline{$d+is$ superconductivity:}}
Next consider a time-reversal symmetry breaking transition in
which the Cooper pair wavefunction in state $X$
acquires a small $s$-wave
component, but with a relative phase factor $\pm \pi/2$ \cite{gabi}:
\begin{equation}
\langle c_{k \uparrow} c_{-k \downarrow} \rangle
= \Delta_0 (\cos k_x - \cos k_y) + i \phi (\cos k_x + \cos k_y).
\label{dsid2}
\end{equation}
The order parameter is again a single real field $\phi$. On
general symmetry grounds, we expect the effective action of the
$\phi$ fluctuations to also have the form (\ref{sI}). However, now
there is an efficient coupling of $\phi$ fluctuations to the nodal
fermions, which is not preempted by momentum conservation. It is
evident from (\ref{dsid2}) that $\phi$ fluctuations Andreev
scatter fermions with momenta $k$ and $-k$. This scattering can
occur between the nodal points $(K,K)$ and $(-K,-K)$ of the
$\Psi_1$ fermions, and similarly for $\Psi_2$; it is
represented by the allowed coupling:
\begin{equation}
{\cal S}_{\Psi\phi} = \int  d^2 x d \tau \Big[ \lambda_0 \phi
\left( \Psi_1^{\dagger} \tau^y \Psi_1 + \Psi_2^{\dagger} \tau^y
\Psi_2 \right) \Big].
\label{dsid4}
\end{equation}
Now, computing the scaling dimension of the coupling $\lambda_0$
under the transformations (\ref{scale}), (\ref{scaleI}), we
observe a crucial difference from the two cases considered so far:
the coupling $\lambda_0$ has dimension $(1-\eta_I)/2 \approx 0.48$,
and is therefore {\em relevant}, and the $\lambda_0=0$
fixed point is unstable. So the critical theory strongly couples
the $\phi$ and $\Psi_{1,2}$ fluctuations, and a complete
understanding requires a more detailed renormalization group
analysis of ${\cal S}_{\Psi} + {\cal S}_{\phi} + {\cal
S}_{\Psi\phi}$. This has been discussed in
Ref.~\cite{vzs} (and for a similar model in a different physical
context in Ref.~\cite{balents}): we will not discuss
this here apart from noting that both the non-linearities,
$\tilde{g}$ and $\lambda_0$ approach fixed-point values, and there
is only one relevant perturbation, ($r-r_c$), at the interacting
critical point. Under these conditions, strong scaling applies,
and the lifetimes of excitations of the $\phi$ and $\Psi_{1,2}$
quanta are of order $\hbar/k_B T$ in the quantum-critical region of
Fig~\ref{fig3} \cite{SY,book}.
So for the case in which $X$ is a $d+is$ superconductor, the
relaxation of the nodal quasiparticles $\Psi_{1,2}$ appears to be
in good accord with experimental observations. However, one
significant discrepancy remains: the $s$-wave order parameter
couples strongly to fermions in all directions, and so will also
couple to the $\Psi_h$ quasiparticles (see Fig~\ref{fig1}).
The gapped $\Psi_h$ quasiparticles will easily radiate many of the
low-energy $\phi$ quanta, and acquire an appreciable width: this is
not in accord with photo-emission experiments in which, as we
noted earlier, the $\Psi_h$ quasiparticles become sharp below
$T_c$.

\paragraph{\underline{$d_{x^2-y^2}+id_{xy}$ superconductivity:}}
Finally, we consider another case in which $X$ breaks time-reversal,
but now by acquiring a small $d_{xy}$ component \cite{did}.
We will see that this
case is very similar to the $d+is$ case discussed above, but it
also succeeds in very simply and naturally resolving the
discrepancy with $\Psi_h$ width we have just mentioned.
The order parameter for this transition, replacing (\ref{dsid2}), is
\begin{equation}
\langle c_{k \uparrow} c_{-k \downarrow} \rangle
= \Delta_0 (\cos k_x - \cos k_y) + i \phi \sin k_x \sin k_y.
\label{dsid3}
\end{equation}
The $\phi$ fluctuations again Andreev scatter fermions between $k$
and $-k$, and their coupling to the $\Psi_{1,2}$ fields has the
form (replacing (\ref{dsid4})):
\begin{equation}
\tilde{\cal S}_{\Psi\phi} = \int  d^2 x d \tau \Big[ \lambda_0 \phi
\left( \Psi_1^{\dagger} \tau^y \Psi_1 - \Psi_2^{\dagger} \tau^y
\Psi_2 \right) \Big].
\label{dsid5}
\end{equation}
Notice that the only difference from (\ref{dsid4}) is the relative
sign of the $\Psi_1$ and $\Psi_2$ terms: this is because the coefficient of $\phi$
in (\ref{dsid3}) changes sign between the two pairs of nodal points, unlike
the case in (\ref{dsid4}).
The consequences of
(\ref{dsid5}) are essentially identical to those of (\ref{dsid4}):
the coupling $\lambda_0$ approaches a fixed point value,
and this leads immediately to a lifetime $\sim \hbar/k_B T$ for
the nodal quasiparticles in the quantum-critical region. Moreover,
the $\Psi_h$ fermions do {\em not} couple to the $\phi$
fluctuations: the coefficient of $\phi$ in (\ref{dsid3}) vanishes
along the line between $(\pi,\pi)$ and $(\pi,0)$ (and also between
$(0,0)$ and $(\pi,0)$ and other symmetry-related lines), and so
the fluctuating $d_{xy}$ component of the pair wavefunction
does not lead to appreciable broadening of the $\Psi_h$
quasiparticles.
Therefore, if the state $X$ is a $d_{x^2-y^2}+id_{xy}$
superconductor, the dynamics of the quantum-critical region are
very naturally in accord with the constraints described at the
beginning of Section~\ref{damp}.

Quite apart from the motivation provided by the above analysis,
there are some appealing independent reasons \cite{did} for suspecting that
the $d$-wave superconductor may be on the verge of an instability
to a $d_{x^2-y^2}+id_{xy}$ state. The structure of the incipient
Fermi surface in Fig~\ref{fig1} indicates that there is significant
second-neighbor hopping on the square lattice. Accompanying this
there should be a corresponding second-neighbor exchange, $J_2$.
Just as the first neighbor exchange, $J_1$, prefers $d_{x^2-y^2}$
pairing, the $J_2$ exchange will induce $d_{xy}$ pairing; the
realistic case with both $J_1$ and $J_2$ non-zero should therefore
prefer the intermediate $d_{x^2-y^2}+id_{xy}$ state, with relative
phase of $\pm \pi/2$ ensuring that a gap opens over the entire
fermion spectrum. Alternatively stated, the $\Psi_h$ fermions are
already strongly paired in the $d_{x^2-y^2}$ state,
while the $\Psi_{1,2}$ fermions are essentially unpaired; the
system will try to lower its energy by pairing the $\Psi_{1,2}$
fermions, and this is most efficiently done by an additional $d_{xy}$
component to the pair wave-function. An additional $s$ component
would also do the job, but has the disadvantage of also deforming
the already optimal pairing of the $\Psi_h$, and so is not as
efficient. Our mean-field calculations \cite{vbs} on models with
$J_1$, $J_2$ both non-zero are consistent with these expectations.
We can therefore identify the coupling $r$ in Fig~\ref{fig3} as
$r \sim J_1/ J_2 $.

It is also important to note that the instability from a
$d_{x^2-y^2}$ superconductor to a $d_{x^2-y^2}+id_{xy}$ superconductor
occurs below a finite value of the coupling $r=r_c$.
This is to be contrasted with pairing
instabilities of a Fermi liquid, which occur at
infinitesimal attraction, and so the analog
of the effective action (\ref{sI}) has a logarithmic dependence upon the order
parameter. However, when the
parent state is a $d_{x^2-y^2}$-wave superconductor, the vanishing
density of states at the Fermi level removes the usual BCS log
divergence, and a finite attraction is
required for further pairing in the
$d_{xy}$ channel. Only such a {\em finite-coupling} quantum
phase transition can be described by an interacting quantum field theory
with hyperscaling properties, and which leads to a $T>0$
quantum-critical region with lifetimes of order $\hbar/k_B T$.

Further tests of the above scenario will be provided by
computations of transport properties, including the optical
conductivities and the Hall coefficient, in the quantum-critical
region of Fig~\ref{fig3} for the case where $X$ is the $d_{x^2-y^2} + i d_{xy}$
superconductor: these are currently in progress.

\acknowledgements

The work reviewed here was performed partly in collaboration with
Chiranjeeb Buragohain \cite{sbv,vbs} and Ying Zhang \cite{vzs}.
We thank Juan-Carlos Campuzano, David Edwards, Behnam Farid, Steve Girvin, Steve
Kivelson, Allan MacDonald, Chetan Nayak, David Pines and especially
Andrey Chubukov for
useful discussions. S.S. is grateful to the participants of the
NATO workshop for many stimulating exchanges.
This research was supported by US NSF Grant No
DMR 96--23181 and by the DFG (VO 794/1-1).


\begin{thebibliography}{99}

\bibitem{vs} M.~Vojta and S.~Sachdev, Phys. Rev. Lett.
{\bf 83}, 3916 (1999).

\bibitem{sbv} S.~Sachdev, C.~Buragohain, and M.~Vojta, Science
{\bf 286}, 2479 (1999).

\bibitem{vbs} M.~Vojta, C.~Buragohain, and S.~Sachdev, Phys. Rev.
B in press,
\href{http://arxiv.org/abs/cond-mat/9912020}
{cond-mat/9912020}.

\bibitem{vzs} M.~Vojta, Y.~Zhang, and S.~Sachdev,
\href{http://arxiv.org/abs/cond-mat/0003163}
{cond-mat/0003163}.

\bibitem{uemura} Y.~J.~Uemura {\em et al.}, Phys. Rev. Lett. {\bf 62},
2317 (1992).

\bibitem{doniach} S.~Doniach and M.~Inui, Phys. Rev. B {\bf 41},
6668 (1990).

\bibitem{nandini} N.~Trivedi and M.~Randeria, Phys. Rev.
Lett. {\bf 75}, 312 (1995); A.~Ghosal {\em et al.},
Phys. Rev. Lett. {\bf 81}, 3940 (1998).

\bibitem{emery} V.~J.~Emery and S.~A.~Kivelson, Nature {\bf 374},
434 (1995); J. Phys. Chem. Solids, {\bf 59}, 1705 (1998).

\bibitem{tvr} A.~Paramekanti, M.~Randeria, T.~V.~Ramakrishnan, and
S.~S.~Mandal,
\href{http://arxiv.org/abs/cond-mat/0002349}
{cond-mat/0002349}.

\bibitem{valla} T.~Valla {\em et al.}, Science {\bf
285}, 2110 (1999).

\bibitem{AbCh} Ar.~Abanov and A.~V.~Chubukov,
\href{http://arxiv.org/abs/cond-mat/0002122}
{cond-mat/0002122};
Ar.~Abanov, A.~V.~Chubukov and J.~Schmalian,
\href{http://arxiv.org/abs/cond-mat/0005163}
{cond-mat/0005163}.

\bibitem{gls} V.~P.~Gusynin, V.~M.~Loktev, and S.~G.~Sharapov,
\href{http://arxiv.org/abs/cond-mat/9811207}
{cond-mat/9811207};
M.~Franz and A.~J.~Millis, Phys. Rev. B {\bf 58}, 14572 (1998);
H.-J.~Kown and A.~T.~Dorsey, Phys. Rev. B {\bf 59}, 6438 (1999).

\bibitem{rossat1} J.~Rossat-Mignod {\em et al.},
Physica C {\bf 185-189}, 86 (1991).

\bibitem{mook} H.~A.~Mook, M.~Yehiraj, G.~Aeppli, T.~E.~Mason, and
T.~Armstrong, Phys. Rev. Lett. {\bf 70}, 3490 (1993).

\bibitem{tony3} H.~F.~Fong, B.~Keimer, F.~Dogan, and I.~A.~Aksay,
Phys. Rev. Lett. {\bf 78}, 713 (1997).

\bibitem{bourges} P.~Bourges in {\it The Gap Symmetry and
Fluctuations in High Temperature Superconductors} ed. J.~Bok,
G.~Deutscher, D.~Pavuna, and S.~A.~Wolf (Plenum, New York, 1998),
\href{http://arxiv.org/abs/cond-mat/9901333}
{cond-mat/9901333}.

\bibitem{he} H.~He {\em et al.},
\href{http://arxiv.org/abs/cond-mat/002013}
{cond-mat/0002013}.

\bibitem{CSY} A.~V.~Chubukov, S.~Sachdev, and J.~Ye,
Phys.  Rev. B {\bf 49}, 11919 (1994).

\bibitem{levin} D.~L.~Liu, Y.~Zha, and K.~Levin, Phys. Rev. Lett.
{\bf 75}, 4130 (1995).

\bibitem{morr} D.~K.~Morr and D.~Pines, Phys. Rev. Lett. {\bf 81},
1086 (1998).

\bibitem{brinck} J.~Brinckmann and P.~A.~Lee, Phys. Rev. Lett.
{\bf 82}, 2915 (1999).

\bibitem{zhang} S.-C.~Zhang, Science {\bf 275}, 1089 (1997).

\bibitem{mopi} P.~Monthoux and D.~Pines, Phys. Rev. B {\bf 49},
4261 (1994); Phys. Rev. B {\bf 50}, 16015 (1994).

\bibitem{AbCh1} Ar.~Abanov and A.~V.~Chubukov, Phys. Rev. Lett.
{\bf 83}, 1652 (2000).

\bibitem{MoCh} D.~K.~Morr and A.~V.~Chubukov, Phys. Rev. Lett.
{\bf 81}, 4716 (1998).

\bibitem{scalapino} D.~J.~Scalapino, Phys. Rep. {\bf 250}, 329
(1995).

\bibitem{carbotte} J.~P.~Carbotte and D.~N.~Basov, Nature, {\bf 401}, 354 (1999).

\bibitem{munzar} D.~Munzar, C.~Bernhard, and M.~Cardona,
Physica C {\bf 312}, 121 (1999).

\bibitem{kami} A.~Kaminski {\em et al.},
\href{http://arxiv.org/abs/cond-mat/0004482}
{cond-mat/0004482};
M.~Eschrig and M.~R.~Norman, 
\href{http://arxiv.org/abs/cond-mat/0005390}
{cond-mat/0005390}.

\bibitem{fong} H.~F.~Fong {\em et al.},
Phys. Rev. Lett. {\bf 82}, 1939 (1999).

\bibitem{mesot} J.~Mesot {\em et al.}, Phys. Rev. Lett. {\bf 83}, 840
(1999).

\bibitem{shen} Z.-X.~Shen and D.~S.~Dessau, Phys. Rep. {\bf 253}, 1 (1995).

\bibitem{campu} J.~C.~Campuzano {\em et al.}, in
{\em The Gap Symmetry and Fluctuations in High-$T_{c}$ Superconductors},
eds. J. Bok {\em et al.} (Plenum, New York, 1998), p. 229.

\bibitem{fed} A.~V.~Fedorov {\em et al.}, Phys. Rev. Lett. {\bf 82}, 2179 (1999).

\bibitem{mesot2} A.~Kaminski {\em et al.}, Phys. Rev. Lett. {\bf 84}, 1788
(2000).

\bibitem{corson} J.~Corson, J.~Orenstein, and J.~N.~Eckstein,
\href{http://arxiv.org/abs/cond-mat/0003243}
{cond-mat/0003243}.

\bibitem{book} S.~Sachdev, {\em Quantum
Phase Transitions}, Cambridge University Press, Cambridge (1999).

\bibitem{CHN} S.~Chakravarty, B.~I.~Halperin, D.~R.~Nelson, {\em Phys.
Rev. B} {\bf 39}, 2344 (1989).

\bibitem{CS} A.~V.~Chubukov and S.~Sachdev, Phys. Rev. Lett.
{\bf 71}, 169 (1993)

\bibitem{sciencereview} S. Sachdev, Science {\bf
288}, 475 (2000).

\bibitem{SY} S.~Sachdev and J.~Ye, Phys. Rev. Lett. {\bf 69}, 2411 (1992).

\bibitem{sigrist1} N.~Nagaosa, A.~Furusaki, M.~Sigrist, and
H.~Fukuyama, J. Phys. Soc. Jpn. {\bf 65}, 3724 (1996).

\bibitem{alloul} H.~Alloul {\em et al.}, Phys. Rev. Lett. {\bf 67}, 3140
(1991).

\bibitem{alloul2a} J.~Bobroff {\em et al.}, Phys. Rev.
Lett. {\bf 79}, 2117 (1997).

\bibitem{julien} M.-H.~Julien {\em et al.},
Phys. Rev. Lett. {\bf 84}, 3422 (2000).

\bibitem{sasha1} A.~V.~Balatsky, M.~I.~Salkola, and A.~Rosengren,
Phys. Rev. B {\bf 51}, 15547 (1995).

\bibitem{kallin} M.~Franz, C.~Kallin, and A.~J.~Berlinsky, Phys.
Rev. B {\bf 54}, 6897 (1996).

\bibitem{fulde} R.~Kilian, S.~Krivenko, G.~Khaliullin, and
P.~Fulde, Phys. Rev. B {\bf 59}, 14432 (1999).

\bibitem{pepin} C.~P\'{e}pin and P.~A.~Lee, Phys. Rev. Lett. {\bf
81}, 2779 (1998);
\href{http://arxiv.org/abs/cond-mat/0002227}
{cond-mat/0002227}.

\bibitem{ting} J-X.~Zhu, D.~N.~Sheng, and C.~S.~Ting
\href{http://arxiv.org/abs/cond-mat/0005266}
{cond-mat/0005266}.

\bibitem{withoff} D.~Withoff and E.~Fradkin, Phys. Rev. Lett. {\bf
64}, 1835 (1990).

\bibitem{CJ} K.~Chen and C.~Jayaprakash,
J. Phys.: Condens. Matter {\bf 7}, L491 (1995).

\bibitem{ingersent} K.~Ingersent, Phys. Rev. B {\bf 54}, 11936
(1996); C.~Gonzalez-Buxton and K.~Ingersent, Phys. Rev. B {\bf
57}, 14254 (1998).

\bibitem{bulla} R.~Bulla, Th.~Pruschke, and A~C.~Hewson,
J. Phys.: Condens. Matter {\bf 9}, 10463 (1997).

\bibitem{SY2} S.~Sachdev and J.~Ye, Phys. Rev. Lett. {\bf 70}, 3339 (1993);
A.~Georges, O.~Parcollet, and S.~Sachdev, Phys. Rev. Lett. to
appear,
\href{http://arxiv.org/abs/cond-mat/9909239}
{cond-mat/9909239}.

\bibitem{si} J.~L.~Smith and Q.~Si,
\href{http://arxiv.org/abs/cond-mat/9705140}
{cond-mat/9705140};
Europhys. Lett. {\bf 45}, 228 (1999).

\bibitem{Sengupta} A.~M.~Sengupta, Phys. Rev. B {\bf 61}, 4041
(2000).

\bibitem{cardybook} J.~L.~Cardy, {\em Scaling and Renormalization in
Statistical Physics}, ch. 7, Cambridge University Press, Cambridge
(1996).

\bibitem{tomo} B. Nachumi {\em et al.}, Phys. Rev. Lett. {\bf
77}, 5421 (1996).

\bibitem{senthil} T.~Senthil and M.~P.~A.~Fisher,
\href{http://arxiv.org/abs/cond-mat/9910224}
{cond-mat/9910224};
\href{http://arxiv.org/abs/cond-mat/9912380}
{cond-mat/9912380};
S.~Sachdev, Phys. Rev. B {\bf 45}, 389 (1992);
Int. J. Mod. Phys. B {\bf 6}, 61 (1992).

\bibitem{affmar} I.~Affleck and J.~B.~Marston, Phys. Rev. B {\bf 37}, 3774
(1988).

\bibitem{schulzoaf} H.~Schulz, Phys. Rev. B {\bf 39}, 2940 (1989).

\bibitem{biq} J.~B.~Marston and I.~Affleck, Phys. Rev. B {\bf 39}, 11538 (1989).

\bibitem{wang} Z.~Wang, G.~Kotliar, X.-F.~Wang, Phys. Rev. B
{\bf 42}, 8690 (1990).

\bibitem{nerses} A.~Nersesyan, Phys. Lett. A {\bf 153}, 49 (1991);
A.~Nersesyan, G.~Japaridze, and I.~Kimeridze, J.~Phys.: Condens.
Matter {\bf 3}, 3353 (1991).

\bibitem{nayak} C.~Nayak,
\href{http://arxiv.org/abs/cond-mat/0001303}
{cond-mat/0001303},
\href{http://arxiv.org/abs/cond-mat/0001428}
{cond-mat/0001428}.

\bibitem{ivanov} D.~A.~Ivanov, P.~A.~Lee, and X.-G.~Wen,
Phys. Rev. Lett. {\bf 84}, 3958 (2000);
X.~G.~Wen and P.~A.~Lee, Phys. Rev. Lett. {\bf 76},
503 (1996); P.~A.~Lee, N.~Nagaosa, T.~K.~Ng, and X.-G.~Wen, Phys.
Rev. B {\bf 57}, 6003 (1998).

\bibitem{gabi} G.~Kotliar, Phys. Rev. B {\bf 37}, 3664 (1988).

\bibitem{balents} L. Balents, M.~P.~A.~Fisher, and C.~Nayak, Int. J. Mod.
Phys. B {\bf 12}, 1033 (1998).

\bibitem{did} D.~S.~Rokhsar, Phys. Rev. Lett {\bf 70}, 493 (1993);
R.~B.~Laughlin, Physica {\bf 243C}, 280 (1994);
R.~B.~Laughlin, Phys. Rev. Lett. {\bf 80}, 5188 (1998);
D.~B.~Bailey, M.~Sigrist, and R.~B.~Laughlin,
Phys. Rev. B {\bf 55}, 15239 (1997);
M.~Sigrist, Prog. Theor. Phys. {\bf 99}, 899 (1998);
D.-H.~Lee, Phys. Rev. B {\bf 60}, 12429 (1999);
A.~V.~Balatsky, Phys. Rev. B {\bf 61}, 6940 (2000);
T.~Maitra,
\href{http://arxiv.org/abs/cond-mat/0002114}
{cond-mat/0002114}.

\end{thebibliography}
\end{document}